# Frequentist prediction intervals for random-effects meta-analysis via confidence-distribution propagation

Hisashi Noma [1,2*] and Guido Schwarzer [3]

[1] Department of Interdisciplinary Statistical Mathematics, The Institute of Statistical Mathematics, Tokyo, Japan

[2] The Graduate Institute for Advanced Studies, The Graduate University for Advanced Studies, Tokyo, Japan

[3] Institute of Medical Biometry and Statistics, Medical Faculty and Medical Center – University of Freiburg, Freiburg, Germany

[*]Corresponding author: Hisashi Noma
Department of Interdisciplinary Statistical Mathematics
The Institute of Statistical Mathematics
10-3 Midori-cho, Tachikawa, Tokyo 190-8562, Japan
TEL: +81-50-5533-8440
e-mail: noma@ism.ac.jp

## Abstract

Prediction intervals are increasingly recommended in random-effects meta-analysis because they describe the range of true effects expected in a future study or setting. Conventional frequentist intervals can have inadequate finite-sample coverage because uncertainty in the between-study variance is not fully propagated. We propose a confidence-distribution propagation method that carries uncertainty through the random-effects hierarchy. The method samples the between-study variance from a confidence distribution obtained by inverting the exact distribution of Cochran's $Q$ and, conditional on each draw, samples the average effect from its corresponding normal confidence distribution before generating a future true effect. Prediction limits are empirical quantiles of the resulting Monte Carlo distribution. Across the scenarios examined, the proposed method improved or maintained coverage relative to existing frequentist intervals, including the Nagashima–Noma–Furukawa confidence-distribution bootstrap, with generally modest increases in expected width. The same Monte Carlo sample also yields confidence intervals for the average effect and heterogeneity measures. The method is implemented in the R package `cdmeta` available at CRAN (https://doi.org/10.32614/CRAN.package.cdmeta).

Key words: confidence distribution; heterogeneity; meta-analysis; prediction interval; random-effects model.

# 1. Introduction

Random-effects meta-analysis is widely used when true effects are expected to vary across studies. [1,2] For many years, the average effect and its confidence interval have been the primary interest in evidence synthesis. However, when substantial between-study heterogeneity is present, a single average effect provides limited information and may be insufficient for interpretation.

Prediction intervals have been increasingly recommended, because they summarize uncertainty in the average effect, between-study heterogeneity, and the variability of the true effect expected in a future study. [3-5] In the frequentist framework, a prediction interval is designed to achieve a specified coverage probability for the true effect in a future comparable study under repeated sampling from the assumed random-effects model. [3,6-8] Commonly used frequentist methods rely on plug-in estimates of the average effect and between-study heterogeneity, often combined with small-sample or degrees-of-freedom adjustments. [6] Although simple and widely implemented, these approaches can have substantial undercoverage when the number of studies is small, particularly because the between-study variance is estimated with considerable uncertainty. [6,7] Bayesian prediction intervals with weakly informative or noninformative priors may also be used, but their frequentist operating characteristics can depend on the prior distribution and may be unsatisfactory in small meta-analyses. [9,10]

Nagashima, Noma and Furukawa [7] proposed a confidence-distribution bootstrap prediction interval, hereafter referred to as the NNF method, that samples the between-study variance $\tau^2$ from its confidence distribution and incorporates uncertainty in the average effect $\mu$ through a $t_{K-1}$-based pivotal term. The method substantially improves finite-sample coverage over conventional plug-in intervals. Our proposal differs in how these uncertainty components are represented and combined. For each sampled value of $\tau^2$, it draws $\mu$ from the corresponding normal confidence distribution and then generates the effect for a future study $\theta_{\text{new}}$ from the random-effects model, yielding an explicit hierarchical propagation scheme. The resulting Monte Carlo sample provides prediction limits together with confidence intervals for $\mu$ and heterogeneity measures. Across the scenarios examined, the proposed method improved or maintained coverage relative to existing frequentist intervals, with a generally modest increase in expected width.

## 2. Prediction intervals in meta-analysis

Consider a meta-analysis of $K$ studies. Let $Y_k$ denote the estimated effect in study $k$, and let $\sigma_k^2$ denote its within-study variance, assumed known or sufficiently well estimated. The normal-normal random-effects model [11,12] is

$$Y_k \mid \theta_k \sim N(\theta_k, \sigma_k^2),$$

$$\theta_k \sim N(\mu, \tau^2),$$

where $\mu$ is the average effect and $\tau^2$ is the between-study variance. Marginally,

$$Y_k \sim N(\mu, \sigma_k^2 + \tau^2).$$

The target of prediction is the true effect in a future study, [3-5]

$$\theta_{\text{new}} \sim N(\mu, \tau^2).$$

If $\mu$ and $\tau^2$ were known, prediction would be straightforward. In practice, both parameters are estimated, and the key difficulty is that $\tau^2$ can be highly uncertain. Because $\tau^2$ directly determines the dispersion of future true effects, treating it as known can produce prediction intervals that are too narrow.

A $100(1-\alpha)\%$ prediction interval is intended to contain $\theta_{\text{new}}$ with probability $1-\alpha$. Higgins, Thompson and Spiegelhalter [3] proposed a commonly used plug-in prediction interval of the form

$$\hat{\mu} \pm t_{K-2,1-\alpha/2}\sqrt{\widehat{\mathrm{SE}}(\hat{\mu})^2 + \hat{\tau}^2},$$

where $t_{K-2,1-\alpha/2}$ is the $1-\alpha/2$ quantile of the $t$-distribution with $K-2$ degrees of freedom. This interval is simple and widely used; it replaces $\mu$ and $\tau^2$ by estimates and uses a $t$-based approximation. Although this produces a simple closed-form interval, it does not propagate the sampling uncertainty of $\hat{\tau}^2$. In particular, $\tau^2$ is often estimated imprecisely when the number of studies is small, and its uncertainty is not fully reflected by plug-in methods. Consequently, the empirical coverage of standard prediction intervals can be substantially below the nominal level. [6-8]

The Higgins–Thompson–Spiegelhalter method is a representative plug-in procedure. [3] Partlett and Riley [6] examined finite-sample performance of confidence and prediction intervals after random-effects meta-analysis, including approaches based on Hartung–Knapp, Sidik–Jonkman, and Kenward–Roger-type adjustments. [13-18] These methods improve inference by modifying variance estimates or degrees of freedom. For example, a Hartung–Knapp-type prediction interval can be written generically as

$$\hat{\mu} \pm t_{\nu,1-\alpha/2}\sqrt{\hat{V}_{\mathrm{HK}} + \hat{\tau}^2},$$

where $\hat{V}_{\mathrm{HK}}$ is a modified variance estimator for the average effect and $\nu$ is a degrees-of-freedom parameter. [6] Such corrections can improve coverage, but they remain approximation-based and depend on the chosen heterogeneity estimator, variance correction, and degrees-of-freedom adjustment.

Nagashima, Noma and Furukawa [7] addressed the plug-in problem more directly by using a confidence distribution for $\tau^2$, derived from the exact distribution of Cochran's heterogeneity statistic. Their method avoids conditioning on a single estimate of $\tau^2$ and substantially improves small-sample coverage. The proposed method retains the confidence-distribution treatment of $\tau^2$ but represents uncertainty in $\mu$ through an explicit draw from its conditional normal confidence distribution rather than through the $t$-based pivotal construction used by NNF.

## 3. Confidence-distribution propagation

A confidence distribution is a data-dependent distribution function defined on the parameter space. It provides a frequentist representation of parameter uncertainty. In appearance, it resembles a Bayesian posterior distribution because it assigns a distribution to an unknown parameter. However, it is constructed from sampling distributions or pivots, and no prior distribution is required. [19,20] Formally, a data-dependent function $H_Y(\psi)$ is a confidence distribution for a scalar parameter $\psi$ if, for each observed dataset, it is a cumulative distribution function on the parameter space and, at the true parameter value $\psi_0$, $H_Y(\psi_0)$ follows a uniform distribution on (0, 1).

The idea is historically connected to Fisher's fiducial inference and Neyman's confidence interval theory. Modern formulations describe a confidence distribution as a "frequentist distribution estimator" of a parameter. [19,20] Intuitively, an ordinary confidence interval gives one interval at one confidence level, such as 95%. A confidence distribution gives a full distribution function from which intervals at any desired level can be obtained. If the confidence distribution is constructed from an exact sampling distribution or an exact pivot, its quantiles have exact or highly accurate frequentist coverage under the assumed model.

This idea is useful for random-effects meta-analysis because uncertainty in $\tau^2$ is the central obstacle in prediction. A single point estimate of $\tau^2$ cannot adequately represent this uncertainty. A confidence distribution for $\tau^2$, by contrast, provides a frequentist distributional summary of plausible heterogeneity values.

It is important to distinguish the levels of exactness involved. The distribution of Cochran's $Q$ statistic used to construct the confidence distribution for $\tau^2$ is exact under the normal-normal model with known within-study variances.[21] Conditional on a fixed value of $\tau^2$, the distribution used for $\mu$ is also exact under the model. The proposed prediction interval is obtained by propagating these components. We do not claim finite-sample exactness of the final marginal prediction interval in all settings; its frequentist operating characteristics are evaluated by simulation.

### *3.1 Proposed frequentist prediction interval*

The proposed interval is constructed by propagating three uncertainty components: heterogeneity uncertainty, mean-effect uncertainty conditional on heterogeneity, and future study variation.

First, we construct a confidence distribution for $\tau^2$. Let the fixed-effect weights be

$$w_k^0 = \frac{1}{\sigma_k^2}.$$

Let

$$W = \mathrm{diag}(w_1^0, \ldots, w_K^0),$$

and let $1$ be the $K$-dimensional vector of ones. Define the weighted centering matrix

$$A = W - \frac{W11^\top W}{1^\top W1}.$$

Define Cochran's heterogeneity statistic as

$$Q = Y^\top AY = \sum_{k=1}^{K} w_k^0 \,(Y_k - \hat{\mu}_{\mathrm{FE}})^2,$$

where

$$\hat{\mu}_{\mathrm{FE}} = \frac{\sum_{k=1}^{K} w_k^0 Y_k}{\sum_{k=1}^{K} w_k^0}.$$

We denote the observed value of $Q$ by $Q_{\mathrm{obs}}$. For a candidate value $t \geq 0$ of $\tau^2$, define

$$\Sigma(t) = \mathrm{diag}(\sigma_1^2 + t, \dots, \sigma_K^2 + t).$$

Under the random-effects model with $\tau^2 = t$, the distribution of $Q$ is

$$Q \mid \tau^2 = t \overset{d}{=} \sum_{j=1}^{K-1} \lambda_j(t) \chi_{1,j}^2$$

where $\chi_{1,1}^2, \dots, \chi_{1,K-1}^2$ are independent chi-square random variables with one degree of freedom, and $\lambda_1(t), \dots, \lambda_{K-1}(t)$ are the positive eigenvalues of

$$\Sigma(t)^{1/2} A \Sigma(t)^{1/2}.$$

Because $A$ removes the mean direction, there are $K-1$ positive eigenvalues, and the distribution does not depend on $\mu$. Let

$$F_Q(q; t) = P\{Q \leq q | \tau^2 = t\}$$

denote the cumulative distribution function of $Q$ under $\tau^2 = t$. The confidence distribution for $\tau^2$ is

$$H_{\tau^2}(t) = 1 - F_Q(Q_{\mathrm{obs}}; t).$$

For fixed $Q_{\mathrm{obs}}$, $F_Q(Q_{\mathrm{obs}}; t)$ is nonincreasing in $t$, so that $H_{\tau^2}(t)$ is nondecreasing in $t$ and defines a confidence distribution on the nonnegative parameter space, with a possible point mass at zero. To generate a Monte Carlo draw, sample $U_b \sim U(0,1)$ and compute the generalized inverse

$$\tau_b^2 = \inf\{t \geq 0 : H_{\tau^2}(t) \geq U_b\}.$$

Because $\tau^2$ is restricted to $[0, \infty)$, this construction allows an atom at zero: in the implementation, $\tau_b^2 = 0$ when $U_b \leq H_{\tau^2}(0)$, and otherwise $H_{\tau^2}$ is inverted over $t > 0$. Repeating this for $b = 1, \dots, B$ gives $\tau_1^2, \dots, \tau_B^2$.

Second, for each sampled $\tau_b^2$, compute

$$w_k(\tau_b^2) = \frac{1}{\sigma_k^2 + \tau_b^2},$$

$$\hat{\mu}(\tau_b^2) = \frac{\sum_{k=1}^{K} w_k\,(\tau_b^2) Y_k}{\sum_{k=1}^{K} w_k\,(\tau_b^2)},$$

and

$$V(\tau_b^2) = \left\{\sum_{k=1}^{K} w_k\,(\tau_b^2)\right\}^{-1}.$$

Then sample

$$\mu_b \sim N\{\hat{\mu}(\tau_b^2), V(\tau_b^2)\}.$$

Third, generate the future true effect by

$$\theta_{\mathrm{new},b} \sim N(\mu_b, \tau_b^2).$$

The proposed $100(1-\alpha)\%$ prediction interval is the empirical $\alpha/2$ and $1-\alpha/2$ quantiles of

$$\theta_{\mathrm{new},1}, \dots, \theta_{\mathrm{new},B}.$$

Marginalizing over the confidence-distribution draw $\mu_b$, the final two sampling steps are equivalently expressed as

$$\theta_{\mathrm{new},b} | \tau_b^2 = t, Y \sim N\{\hat{\mu}(t), t + V(t)\}.$$

Hence, the cumulative distribution function generated by the proposed algorithm can be written as

$$G_{\mathrm{Y}}(x) = \int_{[0,\infty)} \Phi\left\{\frac{x - \hat{\mu}(t)}{\sqrt{t + V(t)}}\right\} dH_{\tau^2}(t)$$

where $\Phi$ denotes the standard normal cumulative distribution function. The explicit draw of $\mu_b$ is not computationally required for construction of the prediction interval, but is retained to separate uncertainty in the average effect from future between-study variation and to provide Monte Carlo draws for inference on $\mu$.

The algorithm can be summarized as follows:

1. Compute $Q_{\text{obs}}$.
2. For each candidate $t$, evaluate $F_Q(Q_{\text{obs}}; t)$ using the weighted chi-square distribution.
3. Construct $H_{\tau^2}(t) = 1 - F_Q(Q_{\text{obs}}; t)$.
4. Generate $\tau_b^2$ from $H_{\tau^2}$.
5. Generate $\mu_b$ from $N\{\hat{\mu}(\tau_b^2), V(\tau_b^2)\}$.
6. Generate $\theta_{\text{new},b}$ from $N(\mu_b, \tau_b^2)$.
7. Use empirical quantiles of $\theta_{\text{new},b}$ as prediction limits.

This construction avoids treating $\tau^2$ as known. Each generated future effect is obtained after first selecting a plausible heterogeneity variance, then a plausible mean effect under that heterogeneity, and finally a plausible future true effect. Figure 1 gives a graphical summary.

The NNF method also accounts for uncertainty in the average effect, but does so through a $t_{K-1}$-based pivotal term in its bootstrap prediction distribution. The proposed method instead represents this component through an explicit draw from the conditional normal confidence distribution of $\mu$, followed by generation of the future true effect. Thus, the distinction is not whether mean-effect uncertainty is included, but how it is represented and propagated. This hierarchical representation also provides the parameter draws used to construct the secondary intervals described in Section 3.2.

### *3.2 Confidence intervals for $\mu$, $\tau^2$, and $I^2$*

The same Monte Carlo sample can also be used to obtain confidence intervals for secondary quantities. For $\mu$, the empirical quantiles of $\mu_1, \dots, \mu_B$ give a confidence interval. This interval incorporates uncertainty in $\tau^2$ through the confidence distribution and conditional uncertainty in $\mu$ through the normal sampling step.

For $\tau^2$, use empirical quantiles of $\tau_1^2, \dots, \tau_B^2$. For $\tau$, compute $\tau_b = \sqrt{\tau_b^2}$ and take empirical quantiles of $\tau_1, \dots, \tau_B$. Since the square-root transformation is monotone, this corresponds to transforming the confidence distribution of $\tau^2$ to the $\tau$ scale.

For $I^2$, choose a typical within-study variance $\sigma_{\text{typ}}^2$. We define

$$I^2 = \frac{\tau^2}{\tau^2 + \sigma_{\text{typ}}^2}.$$

Using fixed-effect weights $w_k^0 = 1/\sigma_k^2$, one may take [22]

$$\sigma_{\text{typ}}^2 = \frac{(K-1)\sum_{k=1}^{K} w_k^0}{\left(\sum_{k=1}^{K} w_k^0\right)^2 - \sum_{k=1}^{K} (w_k^0)^2}.$$

Given $\tau_b^2$, compute

$$I_b^2 = \frac{\tau_b^2}{\tau_b^2 + \sigma_{\text{typ}}^2}.$$

The empirical quantiles of $I_1^2, \dots, I_B^2$ can be used to construct a confidence interval for $I^2$.

These intervals for $\mu$, $\tau^2$, $\tau$, and $I^2$ are useful secondary outputs. The main target of this article, however, is the prediction interval for $\theta_{\text{new}}$.

## 4. Simulation study

We conducted simulation studies to evaluate the frequentist performance of the proposed prediction interval under the normal-normal random-effects model. The simulation settings were chosen to mimic representative meta-analyses of log odds ratios, following scenarios similar to those considered by Brockwell and Gordon. [23] The overall mean effect was fixed at $\mu = 0.50$ corresponding to an odds ratio of 1.65. This value determines only the location of the treatment effects and therefore has no material influence on the evaluation of coverage probabilities or interval widths.

Two sets of scenarios were examined. In the first set, the number of studies was fixed at $K = 5$ or $K = 10$, and the between-study variance $\tau^2$ was varied from 0.01 to 0.20. In the second set, $\tau^2$ was fixed at 0.10 or 0.20, and $K$ was varied from 4 to 20. The within-study variances $\sigma_k^2$ were generated from a scaled chi-square distribution, $0.25\chi_1^2$, and truncated to the interval [0.009, 0.6]. This setting produces heterogeneous study precisions and reflects the imbalance commonly observed in applied meta-analyses.

For each scenario, study-specific true effects were generated from $N(\mu, \tau^2)$, and observed study estimates were generated independently from $N(\theta_k, \sigma_k^2)$. For prediction interval evaluation, an independent future true effect $\theta_{\text{new}}$ was generated from $N(\mu, \tau^2)$ in each replication. We generated 5,000 replicated meta-analyses for each scenario. Empirical coverage probability was calculated as the proportion of replications in which the 95% prediction interval contained $\theta_{\text{new}}$, and expected width was calculated as the average interval length.

The proposed method was compared with existing frequentist prediction intervals implemented in the **`meta`** package. [24] These included the Higgins–Thompson–Spiegelhalter method (HTS), [3] the Veroniki-type method based on a $t$-distribution with $K - 1$ degrees of freedom (V), [25] the Hartung–Knapp method (HK), [13] the Hartung–Knapp method with the Partlett–Riley degrees-of-freedom modification (HK-PR), [6] and the Nagashima–Noma–Furukawa confidence-distribution bootstrap method (NNF). [7] The NNF method was the main comparator because it also uses a confidence distribution for $\tau^2$ and was specifically designed to improve finite-sample performance of prediction intervals. The number of bootstrap or Monte Carlo samples for NNF and the proposed method was set to 5,000.

Figure 2 shows the empirical coverage probabilities of the 95% prediction intervals. Conventional plug-in-type methods showed substantial undercoverage, particularly when the number of studies was small. This was most apparent for $K = 5$ and for small-to-moderate values of $\tau^2$, where uncertainty in the estimated heterogeneity variance was large. The NNF method substantially improved coverage and was generally close to the nominal 95% level. The proposed method generally produced coverage at or above that of NNF across the scenarios examined. In small-sample settings, the proposed method generally shifted coverage upward relative to NNF and reduced residual undercoverage. In some low-heterogeneity settings, this led to mild overcoverage, reflecting the upward shift in coverage under the explicit hierarchical construction.

Figure 3 shows the corresponding expected widths. Methods with poor coverage tended to have shorter intervals. The newly proposed method produced intervals that were slightly wider than NNF in many scenarios, reflecting the alternative uncertainty-propagation scheme and the resulting predictive distribution. However, the increase in width was moderate and was accompanied by improved coverage. Overall, the proposed method improved calibration in the scenarios examined, at the cost of a generally modest increase in expected width.

The frequentist operating characteristics of the confidence interval for $\mu$ were evaluated in a secondary simulation study, with detailed methods and results provided in the Supplementary Material. The proposed interval maintained coverage close to the nominal level and was mildly conservative in some settings with few studies.

## 5. Applications

We applied the proposed method to two illustrative examples with a small number of studies. The first example was the heart failure iron example, based on six randomized trials of intravenous iron therapy for patients with heart failure and iron deficiency. [26] The effect measure was the risk ratio (RR). The second example was the atrial fibrillation VKA example, based on six direct comparisons of vitamin K antagonists versus placebo or control for prevention of stroke in patients with non-valvular atrial fibrillation. [27] The effect measure was the odds ratio (OR). These examples were chosen because both involved only six studies, a setting in which prediction intervals are particularly sensitive to how uncertainty in the heterogeneity variance is handled.

For each example, we applied the same prediction interval methods considered in the simulation study: HTS, V, HK, HK-PR, NNF, and the newly proposed confidence-distribution propagation method. For Monte Carlo-based methods, including NNF and the proposed method, the number of bootstrap samples was set to 5,000. Analyses were conducted on the logarithmic scale, and summary effects, confidence intervals, and prediction intervals were exponentiated for presentation.

Table 1 summarizes the results. In the heart failure iron example, all methods yielded a summary RR below 1, suggesting a beneficial average effect of intravenous iron therapy. However, all prediction intervals included 1, indicating that the true effect in a future study could be close to null or unfavorable despite the beneficial average effect. The conventional prediction intervals were relatively similar across HTS, HK, HK-PR, and V. Both NNF and the proposed method produced wider prediction intervals than the conventional methods. The proposed method yielded a slightly wider prediction interval than NNF, while also providing confidence intervals for the mean effect and heterogeneity measures from the same Monte Carlo sample.

In the atrial fibrillation VKA example, all methods indicated a strong average protective effect of vitamin K antagonists against stroke. The summary ORs were substantially below 1, and the prediction intervals remained below 1 for all methods. Thus, in contrast to the heart failure iron example, the evidence suggested that a protective effect would likely persist in a future comparable study. The proposed method produced a wider prediction interval than the conventional methods and NNF. This difference reflects the alternative uncertainty-propagation scheme and the resulting predictive distribution, rather than the inclusion of uncertainty components omitted by NNF.

Overall, these two examples demonstrate the practical behavior of the proposed method in small meta-analyses. The method gives prediction intervals that are more cautious than conventional plug-in-type intervals and provides a unified set of outputs, including confidence intervals for the average effect $\mu$, $\tau^2$, $\tau$, and $I^2$.

## 6. Discussion

We proposed a novel confidence-distribution propagation method for constructing frequentist prediction intervals in random-effects meta-analysis. The method was motivated by a simple observation: prediction of a future true effect requires propagation of uncertainty in the between-study variance, uncertainty in the average effect conditional on that variance, and future between-study variation. Conventional plug-in methods replace the unknown heterogeneity variance by a single estimate and therefore tend to underestimate the uncertainty. [3,6,7] The proposed method instead samples plausible values of the heterogeneity variance from its confidence distribution, samples the average effect conditionally on each sampled heterogeneity variance, and finally generates future true effects from the random-effects model.

The main advantage of this construction is its transparency. Each generated future effect has a direct interpretation: it is obtained by first selecting a plausible heterogeneity variance, then selecting a plausible average effect under that heterogeneity, and finally generating a future true effect. This sequential propagation of uncertainty provides a natural frequentist approach to prediction intervals while avoiding the specification of prior distributions in Bayesian approaches.

The simulation results showed that the proposed method improved or maintained coverage compared with existing frequentist prediction intervals. Conventional plug-in-type methods frequently showed undercoverage, especially in small meta-analyses. The NNF method substantially improved coverage by combining confidence-distribution sampling of $\tau^2$ with a *t*-distribution based pivotal representation of uncertainty in the average effect. The proposed method uses an explicit hierarchical alternative, drawing $\mu$ conditionally on each sampled value of $\tau^2$ before generating the future true effect. Relative to NNF, this construction generally shifted coverage upward, reducing residual undercoverage in several settings but producing mild overcoverage in some low-heterogeneity, small-sample scenarios. The associated increase in expected width was generally modest.

The proposed intervals were generally slightly wider than those from NNF. This increase should be interpreted together with calibration: in settings where NNF showed residual undercoverage, the additional width was accompanied by coverage closer to the nominal level. Thus, the simulations indicate improved calibration at the cost of a generally modest increase in expected width, rather than uniform superiority on both criteria. In practice, a narrow prediction interval with substantial undercoverage can give an overconfident impression of the likely range of future true effects.

The two applications illustrated the practical implications of the method. In the heart failure iron example, the average effect suggested benefit, but all prediction intervals included the null value, emphasizing that the true effect in a future study may be less certain than the average effect alone implies. In the atrial fibrillation VKA example, the average protective effect was strong and prediction intervals remained below the null across methods, although the proposed method gave a more cautious interval than conventional approaches.

Several limitations should be noted. First, the method is developed under the normal-normal random-effects model and assumes that within-study variances are known or sufficiently well estimated. [1,3,11] This approximation may be less accurate for sparse binary outcomes, rare events, small individual studies, or highly non-normal effect estimators. Second, the confidence distribution for the heterogeneity variance is based on Cochran's $Q$ statistic and the assumed model. If the random-effects distribution is misspecified, or if study-level variances are unstable, the operating characteristics of the method may be affected. Third, although the confidence distribution for $\tau^2$ is derived from the exact distribution of the heterogeneity statistic under the model, we do not claim that the final prediction interval is finite-sample exact. Rather, the proposed method should be viewed as an uncertainty-propagation procedure whose finite-sample performance is assessed empirically.

The method requires numerical evaluation and inversion of the distribution of a weighted sum of chi-square variables. These calculations are computationally feasible for ordinary meta-analyses, although they can become more demanding in large simulation studies or repeated analyses. The method is implemented in the R package **`cdmeta`** available at CRAN (https://doi.org/10.32614/CRAN.package.cdmeta).

Several extensions are possible. The same idea could be adapted to generalized random-effects meta-analysis, rare-event binary outcomes, diagnostic test accuracy meta-analysis, multivariate meta-analysis, and network meta-analysis. Another natural direction is to use the generated future true effects to report probability statements, such as the probability that a future effect exceeds a clinically important threshold. [28] Such summaries would require careful interpretation within the confidence-distribution framework and are left for future work.

In conclusion, the proposed confidence-distribution propagation method provides a simple and effective way to construct frequentist prediction intervals in random-effects meta-analysis. By propagating uncertainty in both the heterogeneity variance and the average effect before generating future true effects, the method improved finite-sample calibration relative to existing approaches across the scenarios examined while remaining easy to implement and interpret.

## Author contributions

Conceptualization: H.N.; Methodology: H.N. and G.S.; Software: H.N. and G.S.; Formal analysis: H.N.; Validation: G.S.; Visualization: H.N.; Writing—original draft: H.N.; Writing—review and editing: H.N. and G.S. Both authors approved the final version of the manuscript.

## Data availability statement

The aggregate datasets used in the applications are part of the R package **`cdmeta`** (https://doi.org/10.32614/CRAN.package.cdmeta). The proposed method is implemented in the R package **`cdmeta`**, with integration into a future version of **`meta`** planned.

## Funding statement

H.N. was supported by Grants-in-Aid for Scientific Research from the Japan Society for the Promotion of Science (grant number: JP23K24811).

## References

[1] Borenstein M, Hedges LV, Higgins JP, Rothstein HR. A basic introduction to fixed-effect and random-effects models for meta-analysis. *Res Synth Methods*. 2010;1(2):97-111.

[2] Egger M, Higgins JP, Smith GD, eds. *Systematic Reviews in Health Research: Meta-Analysis in Context*. 3rd ed. BMJ Books; 2022.

[3] Higgins JPT, Thompson SG, Spiegelhalter DJ. A re-evaluation of random-effects meta-analysis. *J Royal Stat Soc A Stat Soc*. 2009;172(1):137-159.

[4] Riley RD, Higgins JPT, Deeks JJ. Interpretation of random effects meta-analyses. *BMJ*. 2011;342:d549.

[5] IntHout J, Ioannidis JP, Rovers MM, Goeman JJ. Plea for routinely presenting prediction intervals in meta-analysis. *BMJ Open*. 2016;6(7):e010247.

[6] Partlett C, Riley RD. Random effects meta-analysis: Coverage performance of 95% confidence and prediction intervals following REML estimation. *Stat Med*. 2017;36(2):301-317.

[7] Nagashima K, Noma H, Furukawa TA. Prediction intervals for random-effects meta-analysis: A confidence distribution approach. *Stat Methods Med Res*. 2019;28(6):1689-1702.

[8] Noma H, Hamura Y, Sugasawa S, Furukawa TA. Improved methods to construct prediction intervals for network meta-analysis. *Res Synth Methods*. 2023;14(6):794-806.

[9] Hamaguchi Y, Noma H, Nagashima K, Yamada T, Furukawa TA. Frequentist performances of Bayesian prediction intervals for random-effects meta-analysis. *Biom J*. 2021;63(2):394-405.

[10] Noma H. Bayesian estimation and prediction for network meta-analysis with

contrast-based approach. *Int J Biostat*. 2024;20(2):661-676.

[11] DerSimonian R, Laird NM. Meta-analysis in clinical trials. *Control Clin Trials*. 1986;7(3):177-188.

[12] Whitehead A, Whitehead J. A general parametric approach to the meta-analysis of randomised clinical trials. *Stat Med*. 1991;10(11):1665-1677.

[13] Hartung J, Knapp G. A refined method for the meta-analysis of controlled clinical trials with binary outcome. *Stat Med*. 2001;20(24):3875-3889.

[14] Sidik K, Jonkman JN. A simple confidence interval for meta-analysis. *Stat Med*. 2002;21(21):3153-3159.

[15] IntHout J, Ioannidis JP, Borm GF. The Hartung-Knapp-Sidik-Jonkman method for random effects meta-analysis is straightforward and considerably outperforms the standard DerSimonian-Laird method. *BMC Med Res Methodol*. 2014;14:25.

[16] Röver C, Knapp G, Friede T. Hartung-Knapp-Sidik-Jonkman approach and its modification for random-effects meta-analysis with few studies. *BMC Med Res Methodol*. 2015;15:99.

[17] Jackson D, Law M, Rücker G, Schwarzer G. The Hartung-Knapp modification for random-effects meta-analysis: A useful refinement but are there any residual concerns? *Stat Med*. 2017;36(25):3923-3934.

[18] Kenward MG, Roger JH. Small sample inference for fixed effects from restricted maximum likelihood. *Biometrics*. 1997;53(3):983-997.

[19] Xie M, Singh K. Confidence distribution, the frequentist distribution estimator of a parameter: a review. *Int Stat Rev*. 2013;81(1):3-39.

[20] Schweder T, Hjort NL. *Confidence, Likelihood, Probability: Statistical Inference with Confidence Distributions*. Cambridge University Press; 2016.

[21] Biggerstaff BJ, Jackson D. The exact distribution of Cochran's heterogeneity statistic

in one-way random effects meta-analysis. *Stat Med*. 2008;27(29):6093-6110.

[22] Higgins JPT, Thompson SG. Quantifying heterogeneity in a meta-analysis. *Stat Med*. 2002;21(11):1539-1558.

[23] Brockwell SE, Gordon IR. A comparison of statistical methods for meta-analysis. *Stat Med*. 2001;20(6):825-840.

[24] Schwarzer G, Carpenter JR, Rücker G. *Meta-Analysis with R*. Springer International Publishing; 2015.

[25] Veroniki AA, Jackson D, Bender R, et al. Methods to calculate uncertainty in the estimated overall effect size from a random-effects meta-analysis. *Res Synth Methods*. 2019;10(1):23-43.

[26] Anker SD, Karakas M, Mentz RJ, et al. Systematic review and meta-analysis of intravenous iron therapy for patients with heart failure and iron deficiency. *Nat Med*. 2025;31(8):2640-2646.

[27] Dogliotti A, Paolasso E, Giugliano RP. Current and new oral antithrombotics in non-valvular atrial fibrillation: a network meta-analysis of 79 808 patients. *Heart*. 2014;100(5):396-405.

[28] Siemens W, Borenstein M, Evrenoglou T, Meerpohl JJ, Schwarzer G. Beyond prediction intervals in meta-analysis: reporting the expected proportion of comparable studies with clinically relevant benefit or harm. *BMC Med Res Methodol*. 2025;25(1):275.

**Table 1**. Results of the meta-analyses for heart failure iron and atrial fibrillation VKA examples.

| Method | Summary RR/OR (95% CI) | 95% PI | $\tau^2$ (95% CI) | $I^2$ (95% CI) |
|---|---|---|---|---|
| (a) Heart failure iron example | | | | |
| HTS | 0.815 (0.710, 0.935) | (0.584, 1.136) | 0.001 (0.000, 0.285) | 0.344 (0.000, 0.737) |
| HK | 0.815 (0.710, 0.935) | (0.598, 1.110) | 0.001 (0.000, 0.285) | 0.344 (0.000, 0.737) |
| HK-PR | 0.815 (0.710, 0.935) | (0.583, 1.138) | 0.001 (0.000, 0.285) | 0.344 (0.000, 0.737) |
| V | 0.815 (0.710, 0.935) | (0.599, 1.108) | 0.001 (0.000, 0.285) | 0.344 (0.000, 0.737) |
| NNF | 0.815 (0.634, 0.977) | (0.487, 1.186) | 0.009 (0.000, 0.285) | 0.344 (0.000, 0.737) |
| Newly proposed method | 0.804 (0.624, 0.954) | (0.491, 1.203) | 0.036 (0.000, 0.209) | 0.235 (0.000, 0.765) |
| (b) Atrial fibrillation VKA example | | | | |
| HTS | 0.373 (0.271, 0.513) | (0.237, 0.586) | 0.000 (0.000, 0.667) | 0.000 (0.000, 0.746) |
| HK | 0.373 (0.271, 0.513) | (0.263, 0.528) | 0.000 (0.000, 0.667) | 0.000 (0.000, 0.746) |
| HK-PR | 0.373 (0.271, 0.513) | (0.256, 0.543) | 0.000 (0.000, 0.667) | 0.000 (0.000, 0.746) |
| V | 0.373 (0.271, 0.513) | (0.245, 0.566) | 0.000 (0.000, 0.667) | 0.000 (0.000, 0.746) |
| NNF | 0.373 (0.259, 0.538) | (0.193, 0.735) | 0.000 (0.000, 0.667) | 0.000 (0.000, 0.746) |
| Newly proposed method | 0.376 (0.257, 0.562) | (0.195, 0.767) | 0.074 (0.000, 0.593) | 0.126 (0.000, 0.733) |

HTS: Higgins–Thompson–Spiegelhalter; HK: Hartung–Knapp; HK-PR: Hartung–Knapp method with the Partlett–Riley degrees-of-freedom modification; V: Veroniki-type method; NNF: Nagashima–Noma–Furukawa; CI: confidence interval; PI: prediction interval.

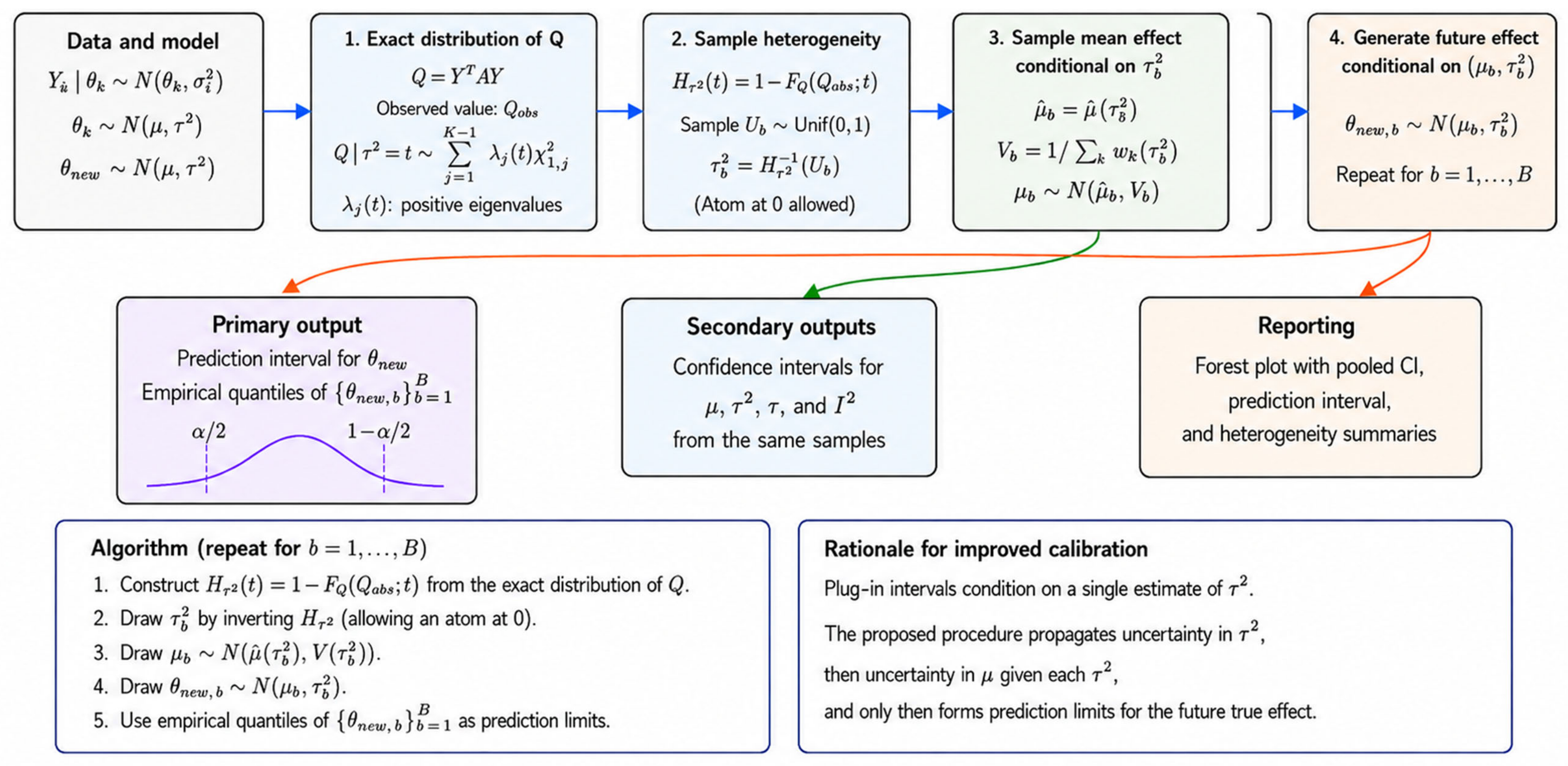

**Figure 1.** Schematic overview of the proposed confidence-distribution propagation method for frequentist prediction intervals.

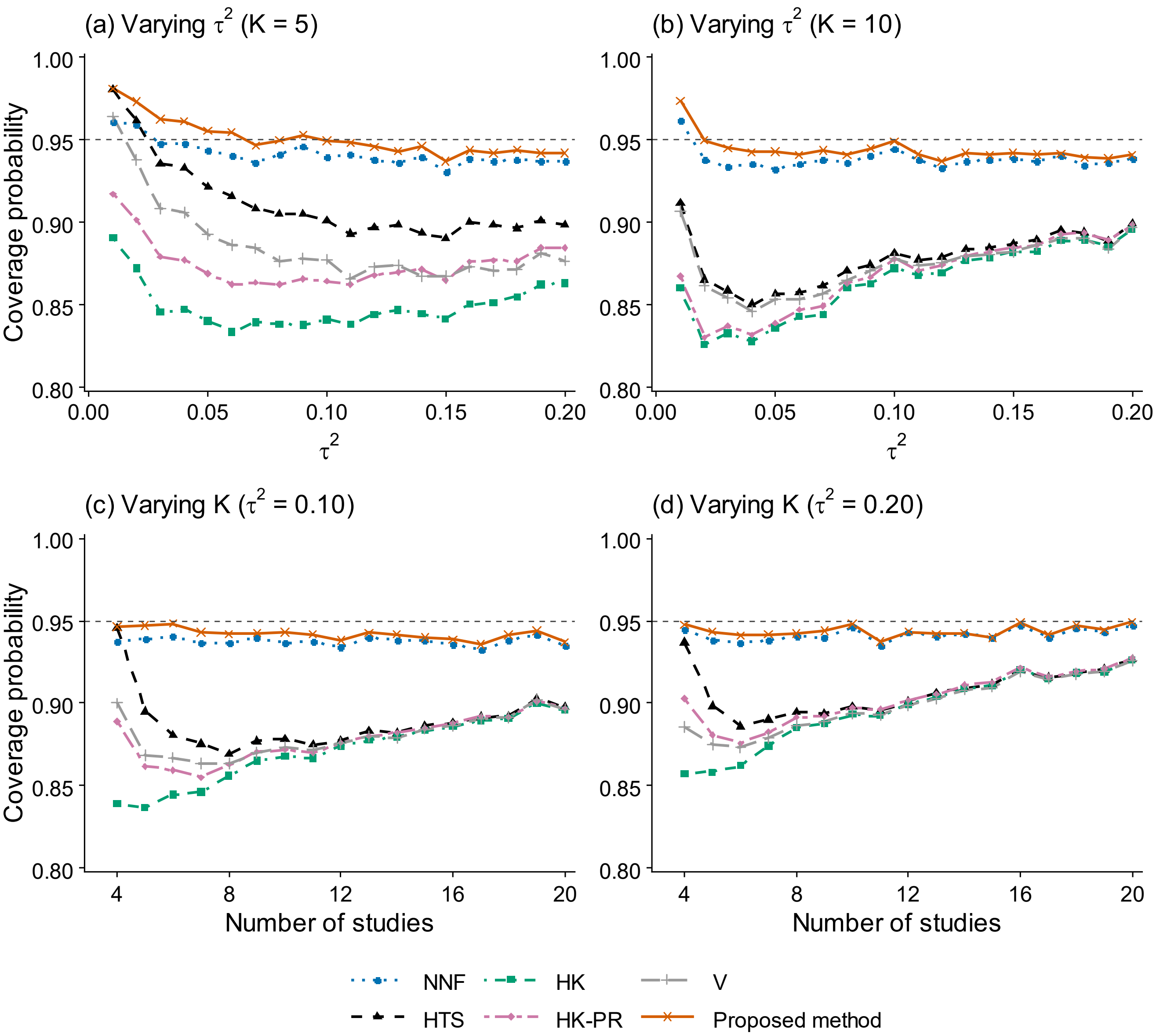

**Figure 2.** Empirical coverage probabilities of 95% prediction intervals across simulation scenarios.

NNF: Nagashima–Noma–Furukawa; HTS: Higgins–Thompson–Spiegelhalter; HK: Hartung-Knapp; HK-PR: Hartung–Knapp method with the Partlett–Riley degrees-of-freedom modification; V: Veroniki-type method.

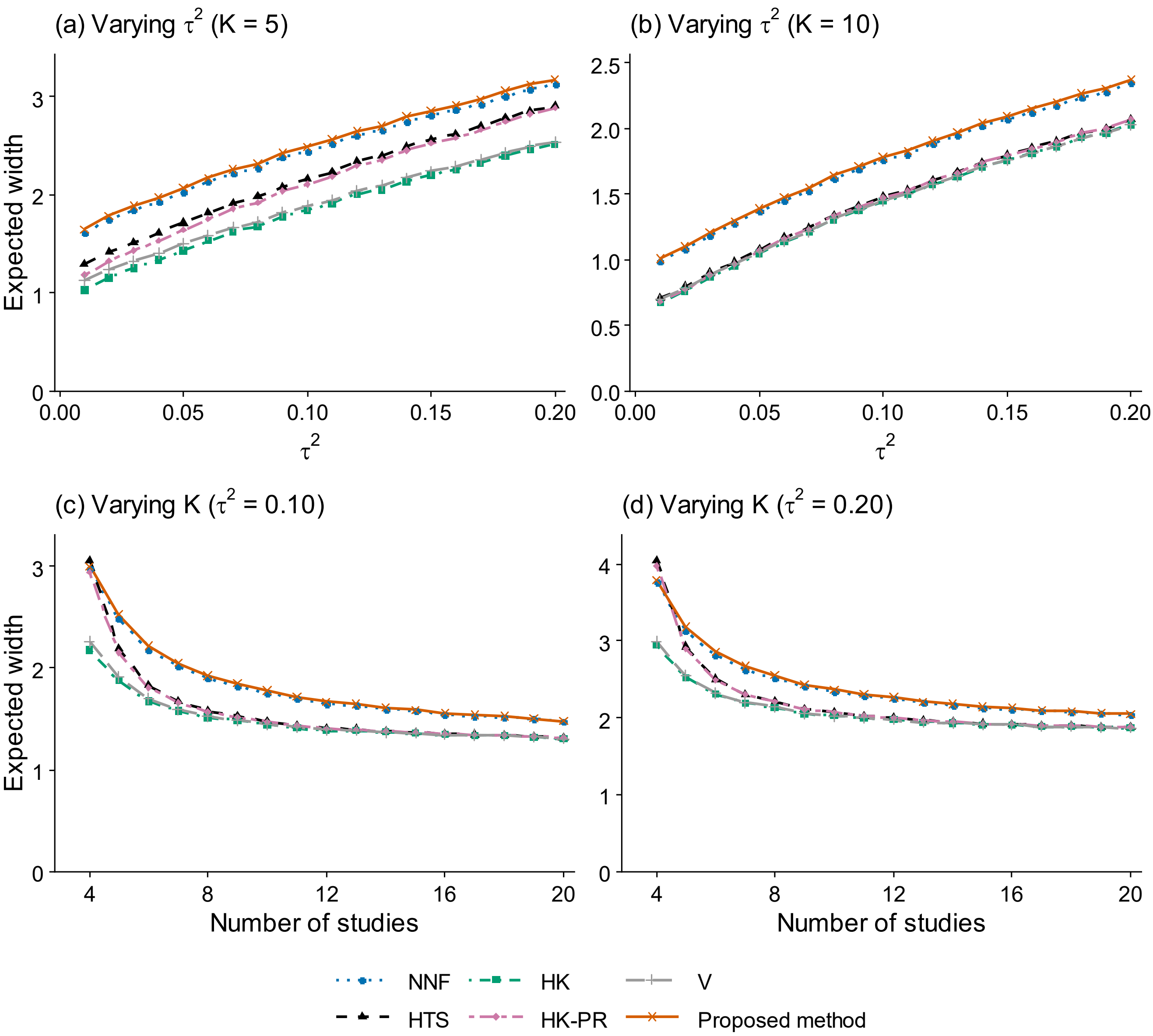

**Figure 3.** Expected widths of 95% prediction intervals across simulation scenarios.

NNF: Nagashima–Noma–Furukawa; HTS: Higgins–Thompson–Spiegelhalter; HK: Hartung-Knapp; HK-PR: Hartung–Knapp method with the Partlett–Riley degrees-of-freedom modification; V: Veroniki-type method.

# Supplementary Material

# Frequentist prediction intervals for random-effects meta-analysis via confidence-distribution propagation

**Hisashi Noma[1,2*] and Guido Schwarzer [3]**

[1] Department of Interdisciplinary Statistical Mathematics, The Institute of Statistical Mathematics, Tokyo, Japan
[2] The Graduate Institute for Advanced Studies, The Graduate University for Advanced Studies, Tokyo, Japan
[3] Institute of Medical Biometry and Statistics, Medical Faculty and Medical Center – University of Freiburg, Freiburg, Germany
* Corresponding author: Hisashi Noma (noma@ism.ac.jp)

## Secondary simulation study of confidence intervals for the average effect

The primary simulation study focused on prediction intervals for the true effect in a future study. Because the proposed confidence-distribution propagation method also generates Monte Carlo draws for the average effect $\mu$, we conducted a secondary simulation study to evaluate the frequentist performance of the resulting confidence interval for $\mu$.

For the proposed method, 5,000 Monte Carlo draws of $\mu$ were generated in each replicated meta-analysis using the procedure described in Section 3.1 of the main text. The 95% confidence interval for $\mu$ was defined by the empirical 2.5th and 97.5th percentiles of these draws, as described in Section 3.2.

The proposed interval was compared with the Hartung–Knapp–Sidik–Jonkman (HKSJ) method (IntHout, Ioannidis and Borm, 2014) and conventional Wald-type confidence intervals based on the DerSimonian–Laird, restricted maximum likelihood, and Paule–Mandel estimators of $\tau^2$ (DerSimonian and Laird, 1986; Paule and Mandel, 1982). For each heterogeneity estimator, the Wald-type interval was constructed using the inverse-variance-weighted estimate of $\mu$ and its conventional normal-theory standard error, with weights determined by the corresponding estimate of $\tau^2$. The HKSJ interval was calculated using the Sidik–Jonkman estimator of $\tau^2$ (Sidik and Jonkman, 2005) and the Hartung–Knapp variance estimator, with critical values from a $t$-distribution with $K - 1$ degrees of freedom (Hartung and Knapp, 2001). No ad hoc variance correction was applied.

Empirical coverage probability was calculated as the proportion of replications in which the 95% confidence interval contained the true value $\mu = 0.50$. Expected width was estimated by the average interval length across replications. With 5,000 replications, the Monte Carlo standard error of an estimated coverage probability of 0.95 was approximately 0.003.

eFigure 1 shows the empirical coverage probabilities of the 95% confidence intervals for $\mu$. The proposed interval maintained coverage close to the nominal level across the scenarios examined and was mildly conservative in several small-sample settings, particularly when $K = 5$. The HKSJ interval also generally achieved coverage close to 95%, although slight undercoverage was observed in some settings with very few studies.

In contrast, the conventional Wald-type intervals based on the DerSimonian–Laird, restricted maximum likelihood, and Paule–Mandel estimators showed substantial undercoverage when the number of studies was small. For fixed $K$, undercoverage generally became more pronounced as $\tau^2$ increased. For fixed $\tau^2$, coverage improved as $K$ increased but generally remained below the nominal level over the range examined. The three Wald-type procedures produced nearly overlapping coverage probabilities, indicating that changing the estimator of $\tau^2$ alone did not resolve the small-sample undercoverage of the conventional normal-theory interval.

eFigure 2 shows the corresponding expected widths. The proposed interval was generally the widest, followed closely by the HKSJ interval. The Wald-type intervals were substantially narrower, particularly when $K$ was small; however, this apparent gain in precision was accompanied by marked undercoverage. Differences between the proposed and HKSJ intervals diminished as the number of studies increased.

Overall, the proposed confidence interval for $\mu$ provided reliable, and sometimes mildly conservative, frequentist coverage as a secondary output of the confidence-distribution propagation procedure. The HKSJ method generally achieved similar calibration with somewhat shorter intervals. The proposed interval should therefore be viewed as a coherent inferential output obtained from the same Monte Carlo sample used for prediction, rather than as a uniformly more efficient alternative to HKSJ when inference on $\mu$ alone is the primary objective.

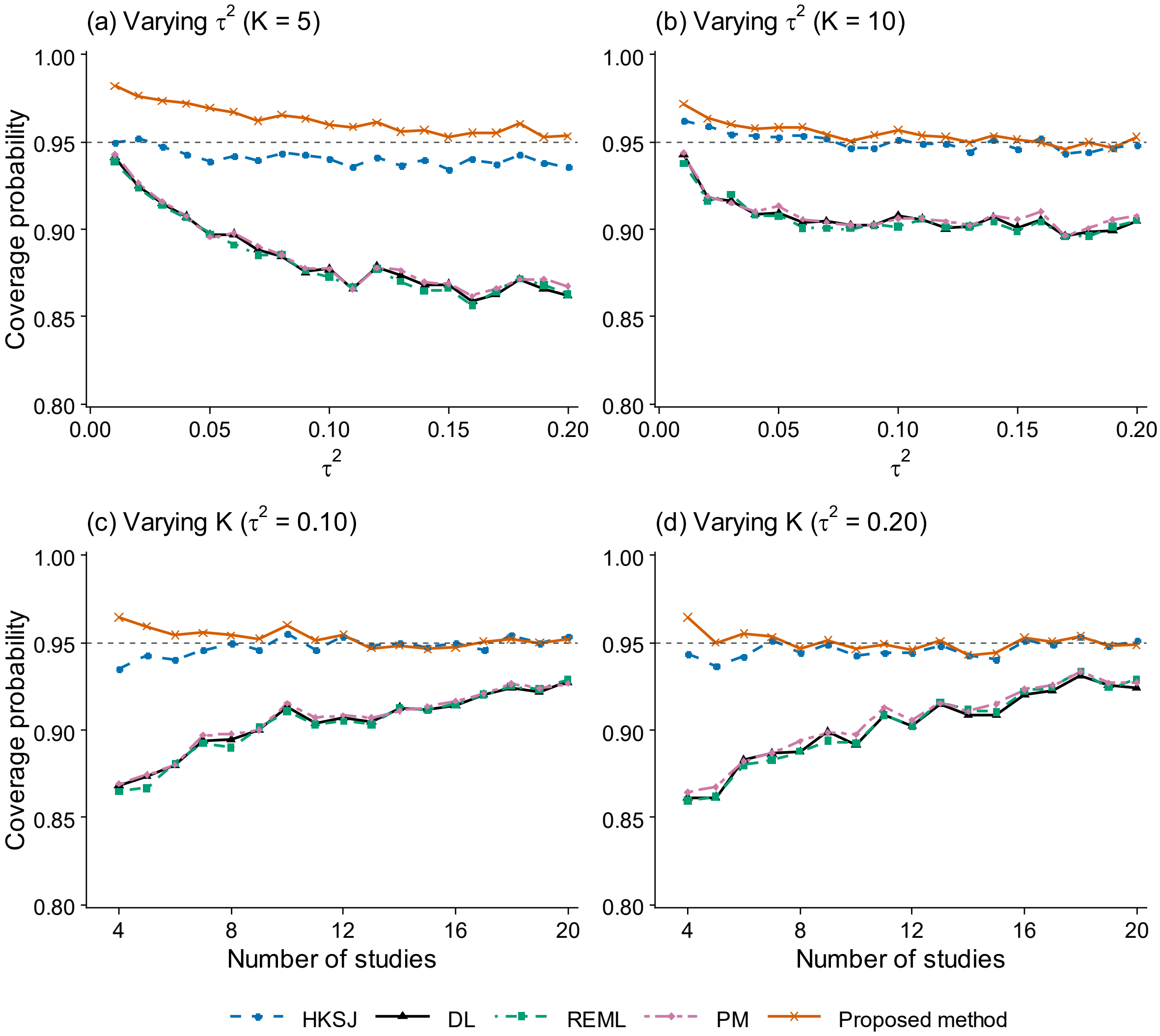

**eFigure 1. Empirical coverage probabilities of nominal 95% confidence intervals for the average effect ($\mu$).**

Panels show results when (a) $\tau^2$ was varied with $K = 5$, (b) $\tau^2$ was varied with $K = 10$, (c) $K$ was varied with $\tau^2 = 0.10$, and (d) $K$ was varied with $\tau^2 = 0.20$. The horizontal reference line indicates the nominal coverage probability of 0.95. Results are based on 5,000 replicated meta-analyses per scenario.

HKSJ: Hartung–Knapp–Sidik–Jonkman confidence interval; Wald–DL: conventional Wald-type confidence interval using the DerSimonian–Laird estimator; Wald–REML: conventional Wald-type confidence interval using restricted maximum likelihood; Wald–PM: conventional Wald-type confidence interval using the Paule–Mandel estimator.

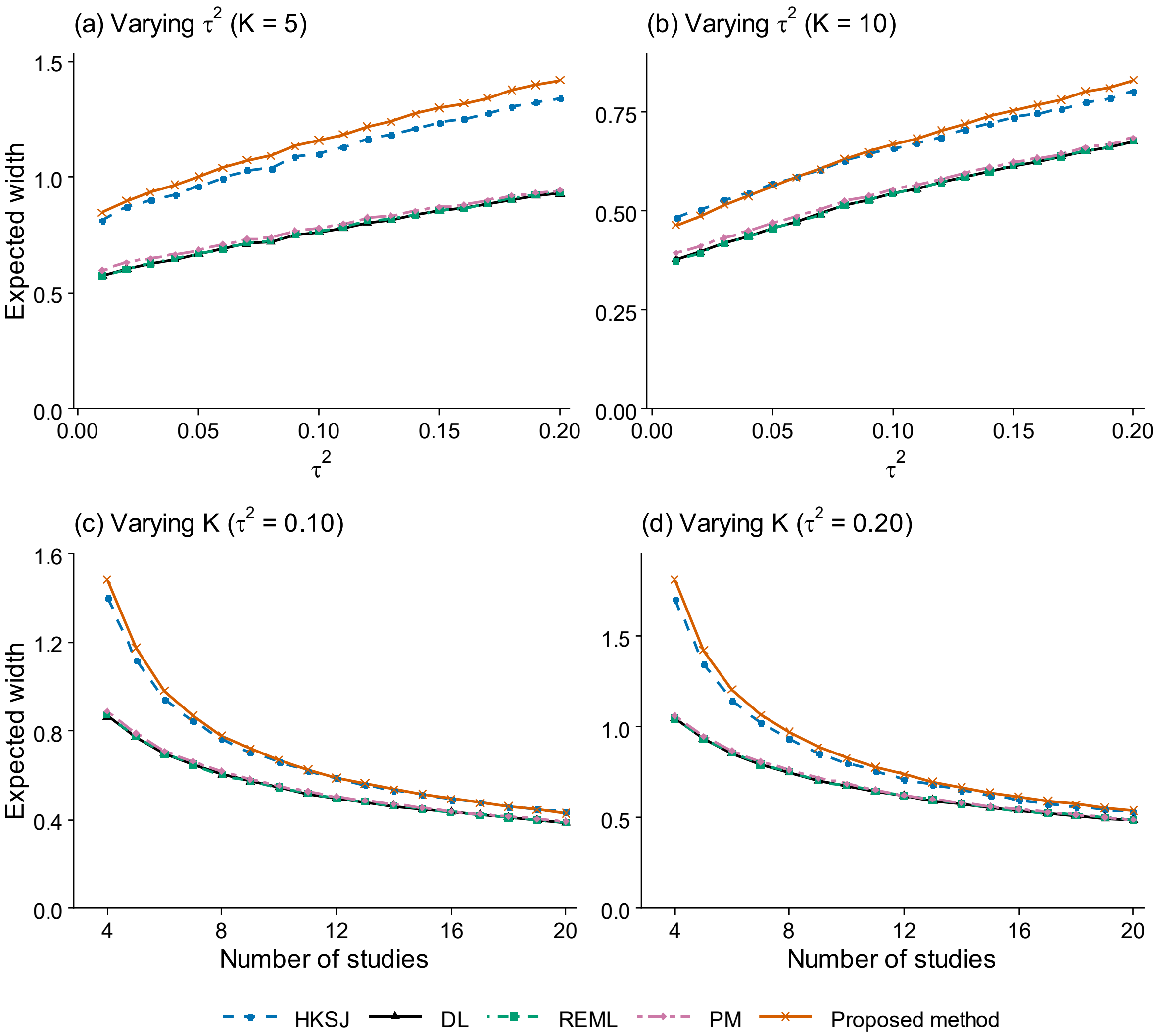

**eFigure 2. Expected widths of nominal 95% confidence intervals for the average effect (*μ*).**

Panels show results when (a) $\tau^2$ was varied with $K = 5$, (b) $\tau^2$ was varied with $K = 10$, (c) $K$ was varied with $\tau^2 = 0.10$, and (d) $K$ was varied with $\tau^2 = 0.20$. Expected width was estimated as the average confidence-interval length over 5,000 replicated meta-analyses per scenario.

HKSJ: Hartung–Knapp–Sidik–Jonkman confidence interval; Wald–DL: conventional Wald-type confidence interval using the DerSimonian–Laird estimator; Wald–REML: conventional Wald-type confidence interval using restricted maximum likelihood; Wald–PM: conventional Wald-type confidence interval using the Paule–Mandel estimator.

## References

DerSimonian R., Laird N. M. (1986). Meta-analysis in clinical trials. *Controlled Clinical Trials*, 7, 177-188.

Hartung J., Knapp G. (2001). A refined method for the meta-analysis of controlled clinical trials with binary outcome. *Statistics in Medicine*, 20, 3875-3889.

IntHout J., Ioannidis J. P., Borm G. F. (2014). The Hartung-Knapp-Sidik-Jonkman method for random effects meta-analysis is straightforward and considerably outperforms the standard DerSimonian-Laird method. *BMC Medical Research Methodology*, 14, 25.

Paule R. C., Mandel J. (1982). Consensus values and weighting factors. *Journal of Research of the National Bureau of Standards*, 87, 377-385.

Sidik K., Jonkman J. N. (2005). Simple heterogeneity variance estimation for meta-analysis. *Journal of the Royal Statistical Society, Series C*, 54, 367-384.